# Pattern Recognition with Magnonic Holographic Memory Device

A. Kozhevnikov[1], F. Gertz[2], Y. Filimonov[1], and A. Khitun[2]

[1] Kotel'nikov Institute of Radioengineering and Electronics of Russian Academy of Sciences, Saratov Branch, Saratov, Russia, 410019

[2] Electrical Engineering Department, University of California - Riverside, Riverside, CA, USA, 92521

**Abstract:**

In this work, we present experimental data demonstrating the possibility of using magnonic holographic devices for pattern recognition. The prototype eight-terminal device consists of a magnetic matrix with micro-antennas placed on the periphery of the matrix to excite and detect spin waves. The principle of operation is based on the effect of spin wave interference, which is similar to the operation of optical holographic devices. Input information is encoded in the phases of the spin waves generated on the several edges of the magnonic matrix, while the output corresponds to the amplitude of the inductive voltage produced by the interfering spin waves on the other side of the matrix. The level of the output voltage depends on the combination of the input phases as well as on the internal structure of the magnonic matrix. Experimental data collected for several magnonic matrixes show the unique output signatures in which maxima and minima correspond to specific input phase patterns. Potentially, magnonic holographic devices may provide a higher storage density compare to the optical counterparts due to a shorter wavelength and compatibility with conventional electronic devices. The challenges and shortcoming of the magnonic holographic devices are also discussed.

Pattern recognition is a widely used procedure, which has multiple applications in text classification, speech recognition, radar processing, and biology[1]. For example, antivirus software installed on a computer system checks the incoming data strings whether it matches one of the data strings (i.e. computer virus) stored in memory. This task requires the exact matching of the input and stored data strings and can be performed by the general type processor within an acceptable time frame.  More complicated, are the problems related to image or speech recognition where the input data may not exactly match the stored data but have some level of similarity [2].  The latter makes real-time processing using a general type processor tremendously difficult or even impossible for large data sets [3]. Holographic data processing is one of the possible solutions, which has been extensively studied in optics during the past five decades[4].  Wave interference is the key mechanism allowing us to reconstruct/recognize a certain pattern even if some part of the data is missing. In contrast to general type logic processors, the increased number of missing bits does not result in the computational overhead for holographic-type devices.   However, to date several technological challenges are currently delaying the practical implementation of optical holographic  devices [5]. One of the key obstacles is on-chip compatibility with conventional electronic integrated circuits it may appear more practically feasible to utilize some other types of waves (e.g. spin waves) for building on-chip holographic co-processors.

The principle of operation magnonic holographic memory (MHM) for data storage and special task data processing is described in Ref. [6].  Spin-wave based realization of optical computing has been also described in Ref.[7].  In brief, MHM devices are comprised of a waveguide matrix with spin wave generating/detecting elements placed on the edges of the waveguides. The matrix consists of a grid of magnetic waveguides connected via cross junctions, where each junction has a magnet placed on the top of the junctions. These magnets act as memory elements holding information encoded in the magnetization state. The read-in and read-out operations of a MHM device are accomplished via spin waves. The first 2-bit MHM prototype comprised of two magnetic cross junctions and two memory magnets has been recently demonstrated[8]. It appeared possible to recognize the four magnetic memory states by the spin wave interference, as the each of the internal magnetic configurations produces a unique output inductive voltage signature. The 2-bit prototype has shown the capabilities of MHM for data storage.  In this letter, we present experimental data illustrating pattern recognition by the eight-terminal MHM device.

The photo and the schematics of the 8-terminal MHM prototype are shown in Figure 1.  The core of the structure is a magnetic matrix comprising a 2×2 grid of magnetic waveguides with magnets placed on top of the waveguide junctions. The waveguides are made of single crystal yttrium iron garnet $Y_3Fe_2(FeO_4)_3$ (YIG) film epitaxially grown on top of a Gadolinium Gallium Garnett ($Gd_3Ga_5O_{12}$) substrate using the liquid-phase transition process.  After the films were grown, micro-patterning was performed by laser ablation using a pulsed infrared laser (λ≈1.03 μm), with a pulse duration of ~256 ns. The YIG matrix has the following dimensions:  the length of the each waveguide is 3 mm; the width is 360 μm; and the YIG film thickness is 3.6 μm. The length of each magnet is 1.1 mm, the width is 360 μm and each has a coercivity of 200-500 Oersted (Oe). There are 8 micro-antennas fabricated on the edges of each waveguide. Antennas were fabricated from a gold wire and mechanically placed directly at the top of the YIG cross. Spin waves were excited by the magnetic field generated by the AC electric current flowing through the antenna(s). The detection of the transmitted spin waves is via the inductive voltage measurements as described in Ref. [9]. The antennas are connected to a Hewlett-Packard 8720A Vector Network Analyzer (VNA) via a number of splitters/combiners with phase shifters and attenuators included in the system. The connection schematics are shown in Figure 1(C). The VNA allowed the S-Parameters of the system to be measured; showing both the amplitude of the signals as well as the phase of both the transmitted and reflected signals. Samples were tested inside a GMW 3472-70 Electromagnet system which allowed the biasing magnetic field to be varied from -1000 Oe to +1000 Oe. The input power is 12.5µW per input port. All experiments are done at room temperature.

The main objective of this work was the demonstration of pattern recognition capabilities of the device. For a data pattern we considered a combination of input phases (e.g. 0, π/4, π/2 ...) generated by the micro-antennas. The output of the device is the inductive voltage detected by the one of the micro-antennas. The input pattern is recognized if the output inductive voltage exceeds some reference value (e.g. 1 mV). For simplicity, we vary only the phases of the waves generated at the three input ports (numbered as 1, 3, and 5 on the connection schematics). The other three antennas (numbered as 2, 4, and 6) generate spin waves of the same constant phase. Hereafter, we take this phase to act as the reference (0) and define the relative phase change at ports 1,3,5 with respect to the reference one. The amplitudes of the spin waves generated by the all six antennas are equalized by attenuators attached to the inputs. The inductive voltage is detected at port 7.

Figure 2 shows experimental data obtained for three magnonic matrixes with different configurations of the junction magnets. The data on X, Y and Z axes in each plot correspond to the Phases 1, 2, and 3, which are the phases of the spin waves generated at ports 1, 3, and 5, respectively. Each plot is a collection of data obtained by the different combinations of three phases, which appears as a cube in the three-dimensional phase space. The level of the output voltage is depicted by the color of the markers, with the red color representing lower-voltage outputs and blue color representing higher-voltage outputs. This is the inductive voltage generated in the micro-antenna at port 7 by the time-varying magnetic flux caused by the propagating spin waves. The direct coupling between the input and output micro-antennas has been extracted by the standard procedure as described in Refs. [10]. There are four plots in Figure 2. The first plot (Fig.2(A)) shows the output for the structure without magnets. Figures 2(B-D) show the output for the specific configurations of micro-magnets. As one can see from Figure 2, each of the magnet configurations produces a unique correlation between the input and the output. The positions of maxima and minima (red and blue color markers) depend on the orientation of the junction magnets.

This correlation between the input spin wave phases and the output voltage amplitude is the base for pattern recognition procedure. We consider input information encoded into the phases of spin waves ($\varphi_1$, $\varphi_2$,... $\varphi_N$, where $N$ is the number of inputs). The MHM device provides an analog voltage output ($V_1$, $V_2$,.. $V_M$, where $M$ is the number of output ports). We classify all possible incoming patterns as "recognized" or "non-recognized" by the level of the inductive voltage produced in the selected output ports (e.g. port 7 in our experiments). For example, the pattern is "recognized" (i.e. stored in MHM) if $V_1<V_r$, where $V_r$ is the reference voltage (e.g. .3mV). All patterns providing output voltage lower than the reference one are considered as "non-recognized" (i.e. not stored in MHM). This final step of pattern recognition requires an analog-to-digital output conversion, which can be accomplished by conventional electronic circuits and is a typical step for light-based techniques [11]. In order to illustrate this procedure, we converted the color plot in Figure 2(A) into a black and white plot in Figure 3(A), where all phase configurations with voltages lower than 0.6 mV are depicted by the black markers, and all outputs higher than 0.6 mV are depicted by white markers. This plot explicitly shows the patterns ($\varphi_1$, $\varphi_2$, $\varphi_3$) stored in the magnetic matrix (magnet configuration is shown in the inset to Fig.2(a)). Next, similar patterns can be also recognized by decreasing/increasing the reference voltage. The Figures 3(B) and 3(C) show the same data as in Figure 3(A) but taken with different reference voltage, Vr=0.4 mV, Vr= 0.5 mV, and Vr=0.6 mV , respectively. The original phase image expands for a higher reference voltage and shrinks for a lower reference voltage. In general, the change of the reference voltage is a powerful tool allowing us to identify similar patterns within a certain Hamming distance.

There are several important observations based on the obtained experimental data we wish to highlight. First, it appears possible to utilize spin waves in a way similar to optical beams for building holograms. Spin wave interference patterns produced by multiple interfering waves are recognized for a relatively long distance (more than 3 millimeters between the excitation and detection ports) at room temperature. In general, the maximum size of magnonic holographic devices is limited by the spin wave coherence length, which is many orders of magnitude shorter than for photons. However, with the current nanometer size fabrication capabilities, coherence length of several millimeters at room temperature is more than enough for building multi-terminal devices. At the same time, the spin wave approach possesses certain technological advantages. The short operating wavelength of spin wave devices promises a significant increase in the data storage density, up to 1Tb/cm$^2$ [6]. Even more importantly, is that spin wave based devices receive input information encoded in voltage and provide voltage at the output, which makes them compatible with conventional CMOS circuitry. Second, the modulation of the propagating spin waves is achieved via the magnetic fields produced by the micro-magnets placed on the top of the spin waveguides. These local magnetic fields significantly affect the amplitude/phase of the propagating spin waves resulting in the change of the output interference pattern (i.e. as shown in Figure 2). The latter opens an intrigue possibility of making re-writable MHM devices, where the position of each magnet is individually controlled (e.g. by the spin-torque devices[12]). It is interesting to note, that spin waves provide an alternative to a magnetoresistance mechanism for read-out, where more than two states of a nanomagnet can be efficiently recognized. Third, the obtained data show a negligible effect to thermal noise and immunity to the structures imperfections. This immunity to the thermal fluctuations can be explained by taking into account that the flicker noise level in ferrite structures usually does not exceed -130 dBm[13].

In order to make MHM devices of practical value, the operating wavelength should be scaled down below 100nm[6]. The main challenge with shortening the operating wavelength is associated with the building of nanometer-scale spin wave generating/detecting elements. The use of micro-antennas is limited due to the fact that the reduction of the antenna's size will lead to a reduction of the inductive voltage. There are other proposed methods of constructing input/output elements including using spin torque oscillators[14], and multi-ferroic elements[15]. Potentially, the utilization of spin torque oscillators makes it possible to scale down the size of the elementary input/output port to several nanometers[16]. Less scalable but more energetically efficient are the two-phase composite multiferroics comprising piezoelectric and magnetostrictive materials[17]. For example, in Ni/PMN-PT synthetic multiferroic reported in Ref. [18], it takes a relatively small electric field of 0.6MV/m has to be applied across the PMN-PT in order to produce 90 degree magnetization rotation in nickel. However, the dynamics of the synthetic multiferroics, especially at the nanometer scale, remains mostly unexplored.

There are many questions on the practical limits for MHM devices development. For example, how many patterns can be stored in one structure? The results presented in this work are obtained for a 2-D magnonic matrix. It would be of great interest to explore the feasibility of building 3-D magnonic structures. However, even a 2-D structure may have an enormous data capacity by exploiting magnets with several thermally stable states. In general, a magnonic N×N matrix with junction magnets having *p* thermally stable states may store as much as $p^{N \times N}$ memory states. The very next question is how many patterns are possible to recognize and how it is related to the operational wavelength? It is reasonable to expect that the minimum size of the junction magnets will be limited by the operational wavelength. On one hand, it is not clear if the magnetic field produced by nanometer scale magnets can provide any prominent effect on propagating spin waves. According to the results of numerical modeling[19], a π-phase shift can be achieved by placing a nano-magnet on top of the magnetic waveguide, though this effect has not been experimental demonstrated yet. Next question is related to the minimum input phase difference which can be recognized at the output. The functional throughput of MHM for pattern recognition can be enhanced by increasing the number of distinguishable phases *k* per input, as the size of the input patters is defined as $k^N$. The higher is the number of distinguishable phases the higher is the functional throughput. Importantly, the increase of the number of input phases does not require any increase of the magnetic matrix or increase the number of output states. It is not required for the output space, with a size of $2^N$ to be the same size as the input space, as the result of computation is simply Yes or No (i.e. there is or there is no such a pattern stored in the memory). At the same time, there is a question on the practically achievable accuracy of the analog output recognition. There is a tradeoff between the accuracy in several millivolts difference in the output and the level of similarity (Hamming distance) one can identify by varying the reference voltage. These are just a few questions which deserve a separate consideration.

In conclusion, we demonstrated an analog device which operation is based on the spin wave interference. A correlation between the phase of the input spin waves and the inductive voltage measured at the output was experimentally observed. It appeared possible to recognize the difference in the output voltage for different combinations of phases in a relatively long device at room temperature. This correlation can be utilized for pattern recognition similar to the procedures developed in optics. Potentially, magnonic holographic devices may provide an advantage over their optical counterparts due to shorter wavelength and compatibility with conventional electronic devices. The main challenge with MHM development is associated with the scaling of the operational wavelength, which, in turn, requires the development of sub-micrometer scale elements for spin wave generation and detection. The development of scalable magnonic holographic devices and their incorporation within integrated circuits may pave the road to the next generations of holographic logic devices aimed not to replace but to complement CMOS in special task data processing.

**Acknowledgments**

This work was supported in part by the FAME Center, one of six centers of STARnet, a Semiconductor Research Corporation program sponsored by MARCO and DARPA and by the National Science Foundation under the NEB2020 Grant ECCS-1124714.

**Figure Captions**

Figure 1. (A) Schematics of the 8-terminal MHM prototype made of YIG with four micro-magnets placed on the top of the cross junctions. The core of the structure is a magnetic matrix comprising a 2×2 grid of magnetic waveguides with magnets placed on top of the waveguide junctions. There 8 micro-antennas fabricated on the edges of each waveguides. (B) Photo of the prototype device packaged. The YIG matrix has the following dimensions: the length of the each waveguide is 3 mm; the width is 360 µm; and the YIG film thickness is 3.6 µm. The length of each magnet is 1.1 mm, the width is 360 µm and each has a coercivity of 200-500 Oersted (Oe). (C) Connection schematics. The antennas are connected to a Hewlett-Packard 8720A Vector Network Analyzer (VNA) via a number of splitters [S], attenuators [A], and phase shifters [P].

Figure 2. Collection of experimental data obtained for four magnonic matrixes: (A) YIG matrix without magnets, (B) matrix with magnets all four magnets directed horizontally; (C) one of the magnets is rotated on 90 degrees; (D) two of four magnets are rotated on 90 degrees. The data on X, Y and Z axes in each plot correspond to the phases of the spin waves generated at ports 1, 3, and 5, respectively. The level of the output voltage in mV is depicted by the color of the markers, with the black color representing lower-voltage outputs and blue color representing higher-voltage outputs.

Figure 3. Experimental data as in Figure 2(A) converted to digital output and plotted for different reference voltages Vr: (A) 0.4 mV, (B) 0.5 mV, and (C) 0.6 mV, respectively. The plots show all phase configurations with output voltages lower than the reference. The phase image expands for higher reference voltage and shrinks for a lower reference voltage.

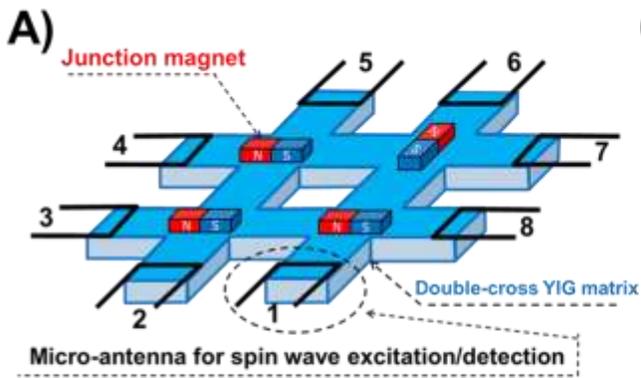
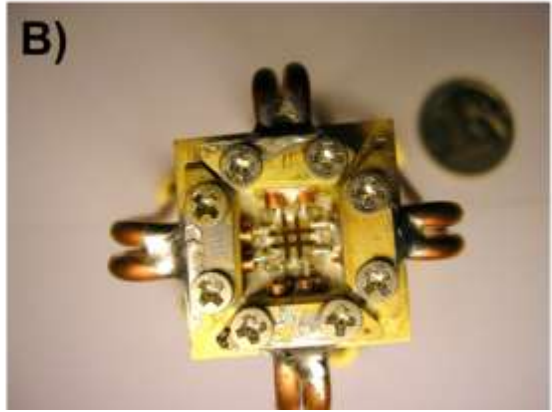
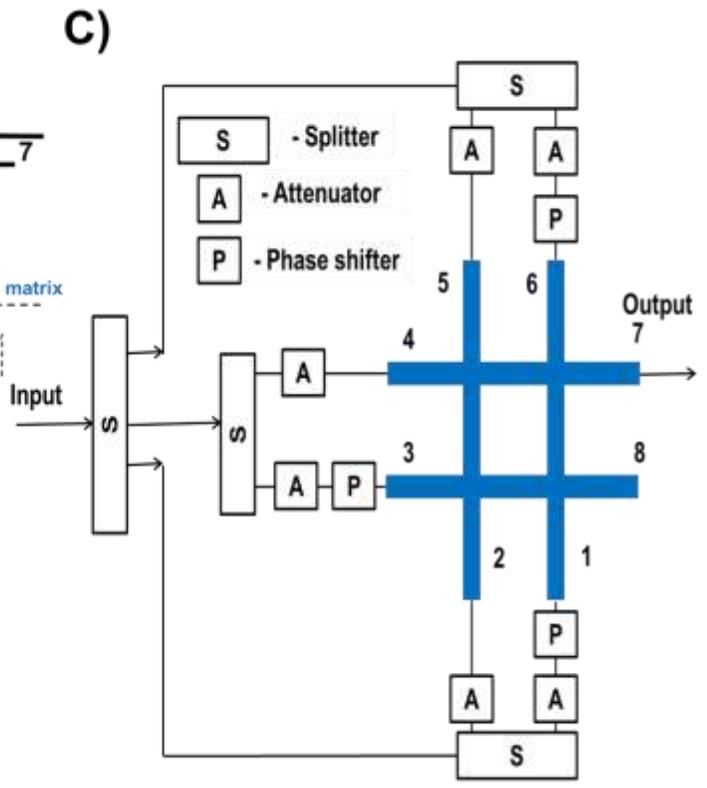

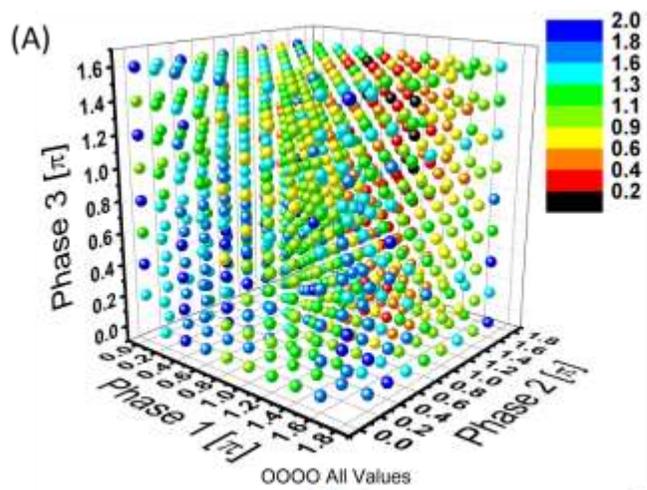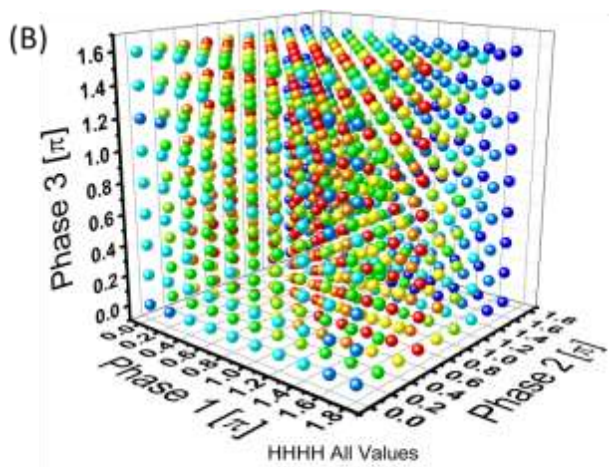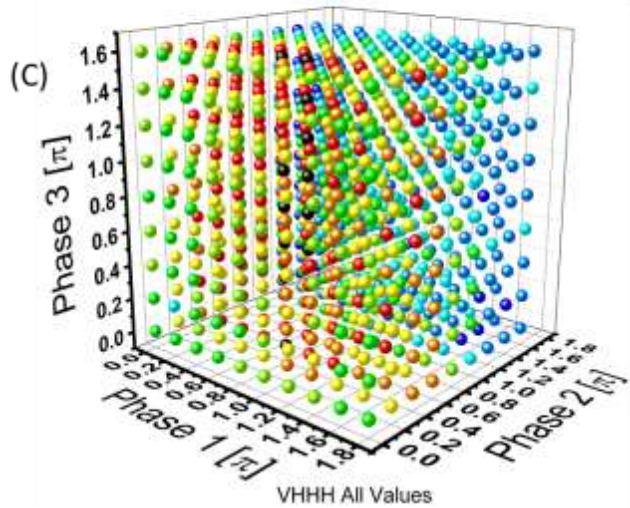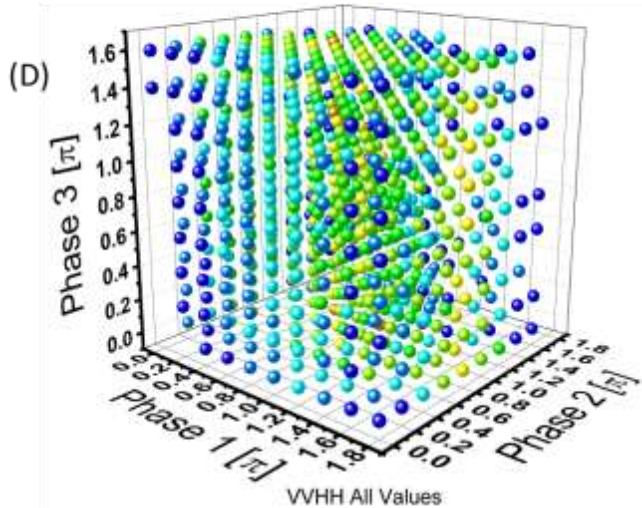

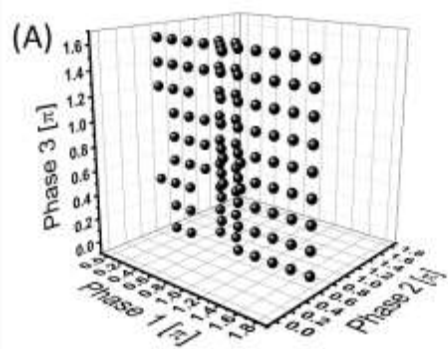 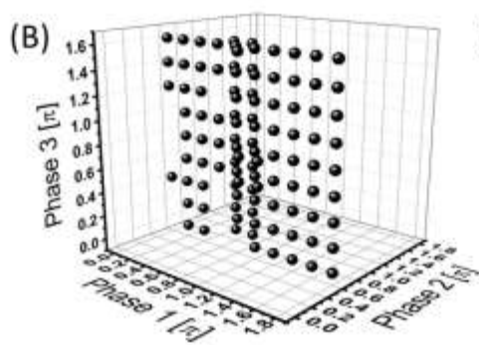 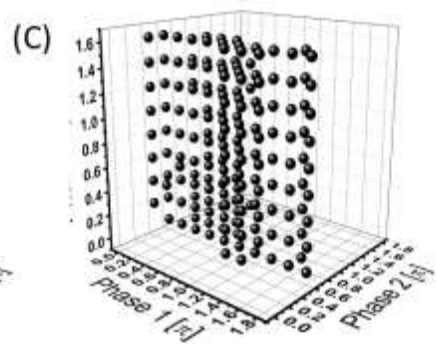